\documentclass[]{aastex631}

\begin{document}

\title{AT\,2023prq: a Classical Nova in the Halo of the Andromeda Galaxy}

\author[0000-0001-6337-6871]{Michael W. Healy-Kalesh}
\affiliation{Astrophysics Research Institute, Liverpool John Moores University \\
IC2 Liverpool Science Park, 146 Brownlow Hill \\
Liverpool, L3 5RF, UK}

\author[0000-0001-8472-1996]{Daniel A. Perley}
\affiliation{Astrophysics Research Institute, Liverpool John Moores University \\
IC2 Liverpool Science Park, 146 Brownlow Hill \\
Liverpool, L3 5RF, UK}

\begin{abstract}
\noindent The classical nova, AT\,2023prq, was discovered on 2023 August 15 and is located at a distance of 46\,kpc from the Andromeda Galaxy (M\,31). Here we report photometry and spectroscopy of the nova. The `very fast' ($t_{2,r^{\prime}}
\sim3.4$\,d) and low luminosity ($M_{r^{\prime}}
\sim-7.6$) nature of the transient along with the helium in its spectra would indicate that AT\,2023prq is a `faint-and-fast' He/N nova. Additionally, at such a large distance from the centre of M\,31, AT\,2023prq is a member of the halo nova population.
\end{abstract}
\keywords{Classical Novae (251)}

\section{Introduction} \label{sec:Introduction}
\noindent Classical novae (CNe) are cataclysmic eruptions emanating from accreting white dwarfs. Hydrogen-rich matter accreted from the donor eventually reaches a critical pressure and undergoes a thermonuclear runaway at the accreted layer's base, resulting in the material being expelled as a nova eruption \citep{1972ApJ...176..169S}.

During eruption, CNe can reach absolute magnitudes up to $M_V{\approx}-10.5$ \citep{2018MNRAS.474.2679A}. Studying CNe in intergalactic regions can aid measurements of intracluster light \citep{2005ApJ...618..692N} and help with tracing populations of intergalactic stars \citep{2006AJ....131.2980S}.

While `hostless' novae, residing far from (but still associated with) a specific nearby galaxy, are expected \citep{2006AJ....131.2980S}, the number known is considerably low: e.g, six in the intracluster regimes of the Fornax Cluster \citep{2005ApJ...618..692N}; four in the halo of M\,31 \citep{1973AN....294..255M} and two in the M\,31 Giant Stellar Stream \citep{2020MNRAS.495.1073D}. Here we report on \object{AT\,2023prq}, a nova found in the halo of M\,31, and one of the most distant novae found from its host.

\section{Observations} \label{sec:Observations}
\subsection{Photometry}
\noindent AT\,2023prq (aka ZTF23aaxzvrr) was discovered by the Zwicky Transient Facility \citep[ZTF;][]{2018ATel11266....1K} on 2023 Aug 15.45 UT with an $r^{\prime}$-band magnitude of 17.13 at $\alpha=1^\mathrm{h}00^\mathrm{m}22^\mathrm{s}\!.28$, $\delta=+42\degr05^\prime13^{\prime\prime}\!\!.57$ \citep[J2000;][]{2023TNSTR1995....1H}. It was observed by the GROWTH-India Telescope \citep[GIT;][]{2022AJ....164...90K} and the Himalayan Chandra Telescope (HCT) on Aug 17.95; the 1.82-m Plaskett Telescope on Aug 18.49 \citep{2023TNSAN.235....1K} and the Asteroid Terrestrial-impact Last Alert System \citep[ATLAS;][]{2018PASP..130f4505T} on Aug 18.51. The last non-detections at the nova's location occurred ${\sim}0.5$\,d prior to discovery: $g{>}18.98$\,mag on Aug 14.45 for ZTF and ${>}19.59$\,mag on Aug 14.51 for ATLAS (`cyan' filter). We followed AT\,2023prq 4.19\,d post-discovery using the $u^{\prime}BVr^{\prime}i^{\prime}$ filters of IO:O on the Liverpool Telescope \citep[LT;][]{2004SPIE.5489..679S} across seven epochs until Aug 31.1.
\begin{figure*}[hb]
\plotone{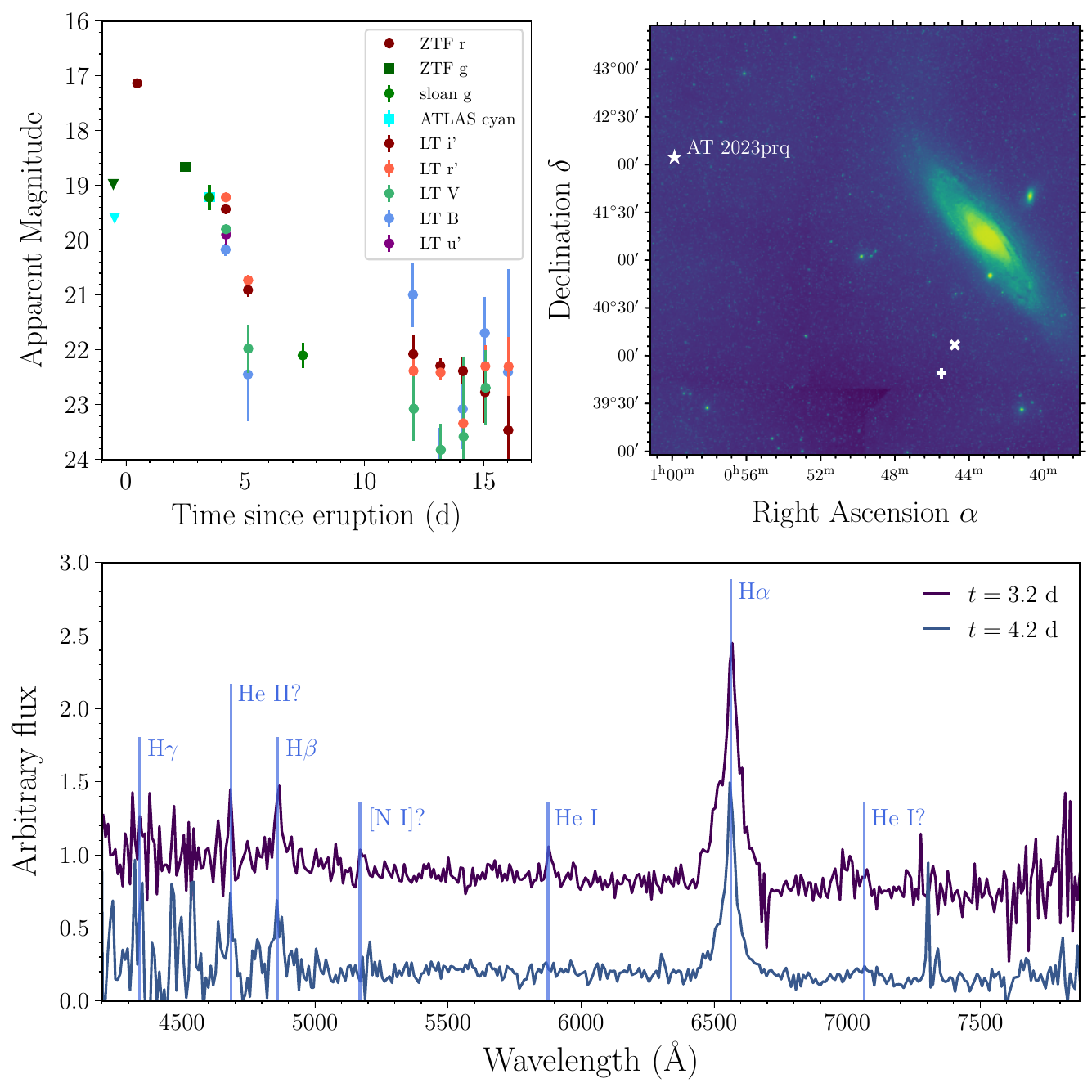}
\caption{{\bf Top left:} Light curve of AT\,2023prq. {\bf Top right:} $4.5^{\circ} \times 4.5^{\circ}$ DSS image of M\,31 and its surroundings. AT\,2023prq is shown (star) with the two tidal stream CNe (AT\,2016dah and AT\,2017fyp) from \citet{2020MNRAS.495.1073D} for comparison. {\bf Bottom:} Spectra of AT\,2023prq with emission lines labelled.
\label{fig}}
\end{figure*}
\subsection{Spectroscopy}
\noindent A spectrum of AT\,2023prq, used for classification, was obtained by \citet{2023TNSCR2017....1P} on Aug 18.19 (3.2\,d post-discovery) using SPRAT on the LT. As reported in \citet{2023ATel16198....1B}, on Aug 19, a spectrum was obtained with the HFOSC on HCT. We obtained optical spectroscopy of AT\,2023prq using SPRAT on the LT on Aug 19.14.

\vspace{12mm}
\section{Results\label{sec:Results}}
\subsection{Photometry}
\noindent Photometry from ZTF, Plaskett Telescope, ATLAS and LT is shown in the top left panel of Figure~\ref{fig}. We take the time of eruption as 2023 Aug $14.98\pm0.47$; this is the midpoint between the last non-detection on Aug 14.51 by ATLAS and discovery on Aug 15.45.

Taking the brightest data point, $r^{\prime} = 17.13$ from ZTF, as the peak apparent magnitude of the nova, a distance to M\,31 of 778\,kpc \citep{1998ApJ...503L.131S} and an extinction of $A_{r^{\prime}}=0.268$, we derive an absolute peak magnitude of $M_{r^{\prime}}{\sim}-7.6$.

Considering the $r^{\prime}$ filter from ZTF and LT $r^{\prime}$ filter as approximately the same to linearly interpolate the data, we estimate decline times of $t_{2,r^{\prime}}=3.4\pm0.5$\,d and $t_{3,r^{\prime}}=4.8\pm0.5$\,d, indicating that AT\,2023prq belongs to `very-fast' speed class. This supports the $t_{2,r^{\prime}}\sim3.1$\,d decline time derived by \citet{2023ATel16198....1B} and their speed classification.

\subsection{Spectroscopy}
\noindent Spectra obtained with LT are shown in the bottom panel of Figure~\ref{fig}. The classification spectrum is labelled $t=3.2$\,d \citep{2023TNSCR2017....1P} and the spectrum obtained on Aug 19.14 is labelled $t=4.2$\,d.

The spectrum taken 3.2\,d post-discovery exhibits Balmer emission lines (H$\alpha$ at 6563\,\AA\ and H$\beta$ at 4861\,\AA) as reported by \citet{2023TNSCR2017....1P}, but also H$\gamma$ at 4340\,\AA\ and He I at 5876\,\AA. There are possibly faint emission lines of He II at 4686\AA, [N I] at 5199\AA\ and He I at 7065\AA. Therefore the spectrum of AT\,2023prq is consistent with belonging to the He/N spectroscopic class. During this epoch, the H$\alpha$ line FWHM is $2100\pm300 \ \text{km} \ \text{s}^{-1}$.

While the spectrum obtained with HCT HFOSC ${\sim}$4\,d after discovery displayed He I at 5876\,\AA\ alongside H$\alpha$ and H$\beta$ emission lines \citep{2023ATel16198....1B}, only Balmer lines are evident within the LT spectrum at a similar epoch (4.2\,d post-discovery). The H$\alpha$ FWHM in the LT spectrum at this epoch decreased to $1700\pm200 \ \text{km} \ \text{s}^{-1}$ \citep[close to the FWHM of $2300 \ \text{km} \ \text{s}^{-1}$ reported by][]{2023ATel16198....1B}.

\section{Discussion\label{sec:Discussion}}
\noindent AT\,2023prq is situated at an angular distance of ${\sim}3.4^{\circ}$ from the centre of M\,31. If the nova is 778\,kpc away, so in the same plane of the sky as the centre of M\,31, this indicates a projected distance from M\,31 of 45.9\,kpc. Whereas, if the nova is in the inclined disk of M\,31, it would be 67.7 kpc from the centre (${\sim}14$ $K$-band scale lengths). AT\,2023prq is not a Galactic nova as it is too faint and does not spectroscopically resemble a dwarf nova. While not possible to rule out, and unlikely, the nova may be a long distance behind M\,31 (up to $\sim$4 Mpc if comparative to the brightest ever nova). Based on the above assumptions, AT\,2023prq is likely to belong to the halo of M\,31, indicating the CN is another example of a `hostless' nova \citep{2006AJ....131.2980S}.

Indeed, if AT\,2023prq is at the same distance as M\,31, then the `very-fast' ($t_{2,r^{\prime}}\sim3.4$\,d) decline time and low absolute magnitude ($M_{r^{\prime}}\sim-7.6$) would place this object in the `faint-and-fast' category of novae \citep{2011ApJ...735...94K,2019MNRAS.486.4334H}. Furthermore, on account of He emission lines and relatively high ejecta velocities derived from the H$\alpha$ emission line, AT\,2023prq, unlike other spectroscopically classified halo novae, belongs to the He/N spectroscopic class.
\vspace{-5mm}

\begin{acknowledgments}
MWH-K acknowledges a PDRA position funded by the UK STFC with grant number ST/S505559/1.
\end{acknowledgments}

\facilities{ATLAS, GIT, HCT, LT, ZTF, 1.82-m Plaskett Telescope}
\software{\texttt{IRAF} \citep{1986SPIE..627..733T}, Starlink \citep{2014ASPC..485..391C}, SAOImageDS9 \citep{2000ascl.soft03002S}}
\vspace{-5mm}
\bibliography{bibliography}{}
\bibliographystyle{aasjournal}

\end{document}